# $^{35}$S BETA IRRADIATION OF A TIN STRIP IN A STATE OF SUPERCONDUCTING GEOMETRICAL METASTABILITY

V. Jeudy, J.I. Collar+, T.A. Girard++, D. Limagne and G. Waysand

Groupe de Physique des Solides (UA 17 CNRS),
Universités Denis Diderot and Pierre et Marie Curie
2, Place Jussieu 75251 Paris Cedex 05, France

## Abstract

We report the first energy loss spectrum obtained with a geometrically metastable type I superconducting tin strip irradiated by the β-emission of $^{35}$S.

+ permanent address: Department of Physics and Astronomy, University of South Carolina, Columbia SC 29208, USA
++ permanent address: Centro de Fisica Nuclear, Universidade de Lisboa, Av. Prof. Gama Pinto 2, 1699 Lisboa, Portugal.

I. Introduction

The development of improved β-spectrometers continues to be of experimental interest since beta decay remains the most sensitive arena to search for the existence of finite neutrino masses [1]. Recently, we reported a preliminary first electron energy loss spectrum with a device based on the phase transition of superheated granules of Type I superconductors, embedded in a dielectric [2]. While these results continue to be investigated, the device suffers the inherent problem of the dielectric presence, which limits the detector resolution due to electron energy losses in it.

Superconducting strips, originally proposed for particle detection more than three decades ago [3], would eliminate this difficulty. Unfortunately, intrinsic metastability is a property of micrometric structures only [4]. The usual method for detecting particles is to bias the strip with a current slightly below the critical current: energy deposition of an incident particle heats the strip, inducing the normal state transition of a slice of the strip which is detected by voltage measurements. In this fashion, strips of In, Sn and Al have been examined under irradiation by 5.5 MeV α-particles [5]; more recently, Nb, g-Al, Sn and W strips have been studied with electrons and 6 keV X-rays [6], Sensitivity and energy resolution are limited most significantly by the strip thickness, which has generally been of 0.2 micron or less [6].

When a superconducting strip is placed in a continuously increasing perpendicular magnetic field, its nucleation is delayed by an energy barrier that prevents spontaneous flux penetration into the volume. As suggested by J.R. Clem et al. [7], this metastability is determined by the sample geometry (width, thickness and length) and is not intrinsic to the superconducting properties of the material as in the case of the superheating and supercooling observed with micrometric granules. We have recently demonstrated this geometrical metastability in strips of relatively large thickness (20 micron), of both Type I & II superconductors [8]; over a large range of applied magnetic field values, the transition is achieved through the penetration of multiquantum flux bundles, nucleated on the strip edges. Here we show that the rupture of the metastability can be achieved by thermal nucleation resulting from energy deposition of incident particles. First results obtained with a $^{35}$S β source show that the amplitude of the detected signal is proportional to the energy deposited.

2. Experimental description

The strip was cut from a 99.9% pure [9], annealed, pinhole-free, 20 ± 1 µm thick ($L_y$) polycristalline tin foil. Its width ($L_x$) and length ($L_z$) were 700 ± 45 µm and 2.45 ± 0.10 cm, respectively. The β source was obtained by evaporating $^{35}$S (100% β⁻ decay, $Q_{\beta^-}$ = 167.5 keV) into a 20 µm thick smoking paper. The total source activity is 4 mCi, distributed over the 3 cm$^2$ paper surface.

Experiments were performed in a single shot $^3$He refrigerator [10]. The $^3$He bath was maintained under reduced pressure by charcoal pumps, and cooled a copper plate to T= 390 ± 10 mK. The prototype detector consisted of a multi-layered sandwich: the source was placed against the strip; a 1.5 µm thick mylar sheet was intercalated between the strip and a copper pick-up coil, supported by an epoxy board; another mylar sheet was placed between the source and the cold plate. The sandwich was mechanically fastened to the cold plate by screws in an epoxy board, as shown in the insert of Fig. 1.

The flux penetration was detected by the same fast-pulse acquisition system employed with metastable superconducting granules detectors [2,11]. This system synchronises the magnetic step-rise with the opening of a gate during which flux pulses are detected. The nucleation of flux bundles creates discontinuities in the flux cut by the detecting loop. This loop is connected via two transformers to a LeCroy HQV810 based pre-amplifier. The detecting chain is sensitive to flux bundles of a few micrometers in diameter [11]. Only irreversible flux entry is detected: reversible flux changes are outside of the bandwidth of the pre-amplifier. The input signal is then shaped with a fast LeCroy amplifier and discriminated with a LeCroy MVL407 ultrafast voltage comparator. For the amplitude analysis, the amplifier output was branched in parallel to the input of a 10 bit multichannel analyser (LeCroy qVt 3001).

The magnetic field, applied perpendicularly to the strip, was raised from zero at a rate of 16 G / s. The ramping was stopped for 10 s at a pause field ($H_{pause}$) of 65 G, chosen just above the onset of the irreversible penetration where the highest number of counts due to irradiation was detected. The magnetic field ramping was then resumed to 500 G, well above the thermodynamical critical field at the operating temperature, $H_c$(390mK), and returned to zero.

A typical signal acquired during the pause is shown in Fig. 1. Two decay processes are superimposed: the exponential decay time of the applied magnetic field damping in the coils is 100 ms; the second effect is due to the penetration of flux tubes into the superconducting strip, induced by the electron energy deposition. The time constant of the irradiation exponential decay is about 1.79 s; after 8 s, no particle is detected: the detector is saturated and must be reset for new particle detection.

3. Results and Discussion

Fig. 2 shows the pulse-height spectrum of the detector output. To obtain the experimental energy absorption spectrum, the procedure described above was repeated 30 times to improve the statistics and the 1024 channels were grouped by 16. The experimental curve (Fig. 2: •) exhibits a hump centred around channel # 300.

A Monte Carlo simulation was performed to obtain an estimate of the electron energy loss in the strip and to identify the origin of the hump. Fig. 2 shows a preliminary comparison between the pulse-high spectrum and the simulation of the electron energy deposition in the strip. This includes energy losses at the source, Molière straggling [12] (Coulomb scattering of low energy electrons) and backscattering at the source-detector interface using the model of Archard [13]. Energy losses at both source and strip are computed via the Bethe-Bloch equation. In order to calibrate the pulse-height spectrum, the region around the last occupied channel (# 971) was fitted with a linear function. The experimental end-point was found at channel # 949±24, with the uncertainty arising from the choice of fitted region. This end-point was made to correspond to the Q-value of the β emission spectrum (167.5 keV). With this, the excess of counts at ch# ≈ 300 and of deposited energy at ≈ 45 keV coincide within a 2.5 % precision. This excess comes from the fact that electrons emitted with energies higher than ~ 70 keV can cross the strip thickness (20 μm) without losing all their energy. This together with backscattering promote partial energy deposition. We found a dependence of the simulated energy deposition on the choice of theoretical model of Molière straggling employed. The backscattering theory of Archard gave the best agreement between experiment and simulation. The simulation does not include any thermodynamics and is therefore a yet incomplete description of the strip's response function; an improved, much thinner $^{109}$Cd source (monochromatic electron emission) will facilitate its completion.

The hump is observed to occur at the same position in both spectra, indicating the linearity of the signal response. Since the time variation of the magnetic flux in the coil, $d\phi/dt$, is time-integrated by the electronics, the pulse-height recorded is proportional to the flux contained in the thermally nucleated normal tube, i.e., pulse-height ~ $\int (d\phi/dt) dt$ ~ $\phi = H_c(T) S_{tube}$, where $S_{tube}$ is the surface area of the flux tube. The observed linearity of the spectrum then implies that the nucleated volume $L_y S_{tube}$ (where $L_y$ is thickness of the strip) must be proportional to the energy deposited by a particle at a given temperature:

$$\frac{E_{dep}}{L_y S_{tube}} = \text{constant.}$$

4. Conclusions

Geometric metastability can be broken by thermal nucleation. The energy deposition of electrons, emitted by a $^{35}$S source, induces the irreversible penetration of multiquantum flux tubes. This first experimental absorption spectrum and simulation are in good agreement, indicating a linear response of the detector to the incident particles.

We intend to explore the $\beta^-$ decay of $^{187}$Re ( $Q_{\beta^-}$= 2.67 keV, $T_{1/2}$= 5x10$^{10}$ yr, 62.6% natural isotopic abundance) using a rhenium thin strip. The decay is internal to the detector, with the consequent improvement in resolution. For the same reason, the molecular and atomic binding corrections that dominate conventional $^3$H neutrino-mass searches, do not apply [14]. The resolution losses intrinsic to cryogenic bolometers and arising from thermal coupling to the detection chain may be minimized in such a device, where the only bias is through the applied magnetic field. We expect to largely reduce the electronic noise threshold by using a cooled FET or a SQUID-amplifier. This search would be an alternative to current $^3$H experiments and may cast some light on their anomalous results [15], specifically on the unphysical negative neutrino-mass values so far obtained.

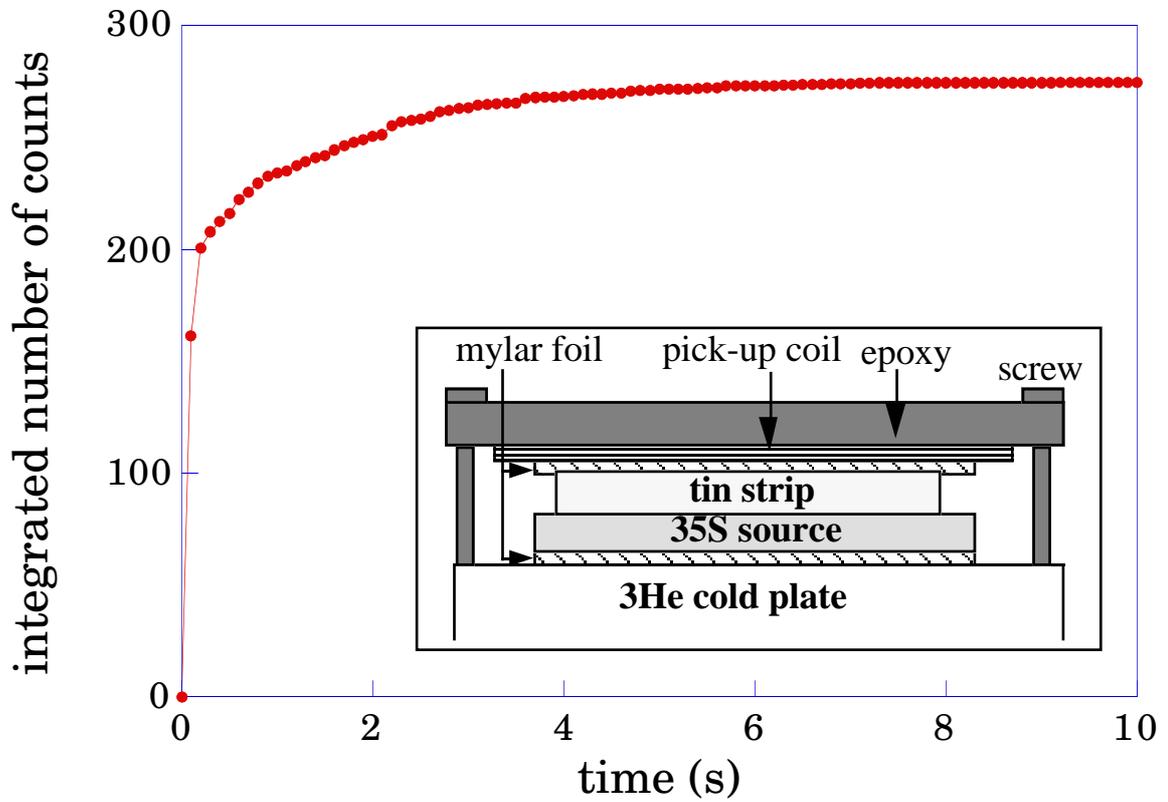

Fig. 1.

Integrated number of detected pulses during a 10 s pause at 65 G (average over 14 runs). The efficiency of the irradiation detection decreases exponentially. Insert: experimental set-up.

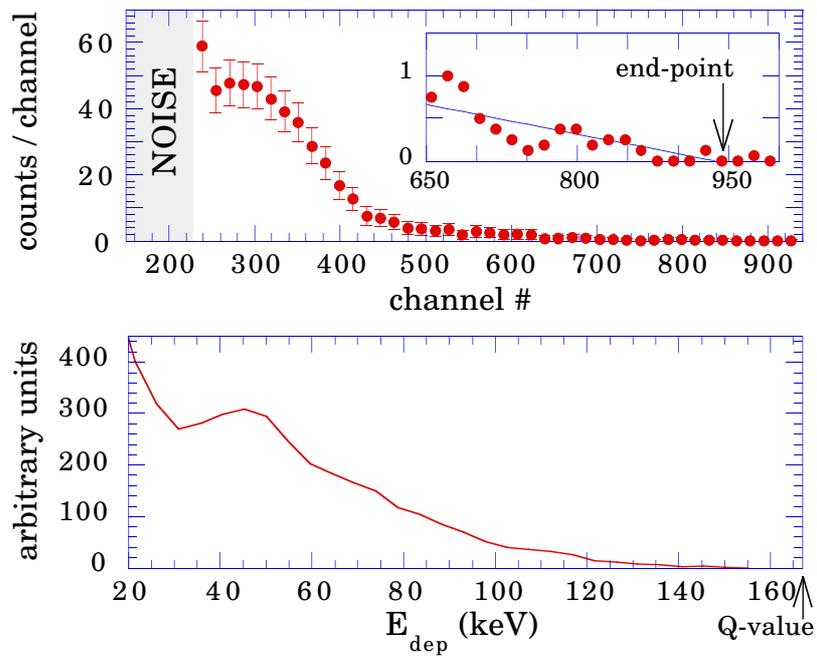

Fig. 2.

Top figure: sum, over 30 runs, of the number of counts recorded during the 10 s pauses, vs. channel #. The electronic noise exceeds the range of the graph. The insert is a magnification of the end-point region, showing the best linear fit. Bottom figure: the simulated absorption spectrum calibrated so that the Q-value coincides with the end-point of the experimental spectrum.